\documentclass{statsoc}

\usepackage[a4paper]{geometry}
\usepackage{graphicx}
\usepackage[textwidth=8em,textsize=small]{todonotes}
\usepackage{amsmath}
\usepackage{natbib}
\RequirePackage{graphicx}
\usepackage{amsmath,amsfonts,amssymb}

\usepackage{float}
\usepackage{xcolor}

\newtheorem{theorem}{Theorem}[section]
\newtheorem{lemma}[theorem]{Lemma}
\newtheorem{proposition}[theorem]{Proposition}

\def\nat{{\mathbb N}}

\def\spX{\mathcal{X}}

\def\CB{\mathbb{D}}
\def\prob{{\mathbb P}}

\def\copula{\widehat{\mathbb{D}}}
\newcommand{\norm}[1]{\lVert{#1}\rVert}

\def\real{{\mathbb R}}

\newcommand{\PP}{\mathbb{P}}
\def\conAS{\stackrel{\mathrm{a.s.}}{\longrightarrow}}
\def\conD{\Rightarrow}
\def\conW{\leadsto}
\def\spX{\mathcal{X}}
\newcommand{\bx}{{\boldsymbol{x}}}
\newcommand{\diff}{\mathrm{d}}
\newcommand{\bu}{{\boldsymbol{u}}}
\def\bv{{\bf v}}

\title[RFA and dependence]{Non-parametric multimodel Regional Frequency Analysis applied to climate change detection and attribution}

\author[Le Gall {\it et al.}]{Philomène Le Gall}
\address{Univ. Grenoble Alpes, CNRS, IRD, Grenoble INP, IGE, F-38000 Grenoble, France}
\email{philomene.le-gall@univ-grenoble-alpes.fr}
\author{Anne-Catherine Favre}
\address{Univ. Grenoble Alpes, CNRS, IRD, Grenoble INP, IGE, F-38000 Grenoble, France}
\author{Philippe Naveau}
\address{Laboratoire des Sciences du Climat et de l'Environnement, ESTIMR, CNRS-CEA-UVSQ, Gif-sur-Yvette, France}
\author{Alexandre Tuel}
\address{Institute of Geography and Oeschger Centre for Climate Change Research, University of Bern, Switzerland}

\begin{document}
\begin{abstract}
A recurrent question in climate risk analysis is determining how climate change will affect
heavy precipitation patterns.
Dividing the globe into homogeneous sub-regions should improve the modelling of heavy precipitation by inferring common regional distributional parameters.
In addition, in the detection and attribution (D\&A) field, biases due to model errors in global climate models (GCMs) should be considered to attribute the anthropogenic forcing effect.
Within this D\&A context, we propose an efficient clustering algorithm that, compared to classical regional frequency analysis (RFA) techniques,  is covariate-free and accounts for dependence.
It is based on a new non-parametric dissimilarity  that combines both the RFA constraint and the pairwise dependence.
We derive asymptotic properties of our dissimilarity estimator, and we interpret it for generalised extreme value distributed pairs.

As a D\&A application, we cluster annual daily precipitation maxima of 16 GCMs from the coupled model intercomparison project.
We combine the climatologically consistent subregions identified for all GCMs.
This improves the spatial clusters coherence  and outperforms methods either based on margins or on dependence.
Finally, by comparing the natural forcings partition  with the  one with all forcings, we assess the impact of anthropogenic forcing on precipitation extreme patterns.

\end{abstract}

\section{Introduction}\label{sec: intro}


Since the early 19$^\text{th}$ century, fossil fuels-based human activities have become one of the major forces of ecosystem and climate change, defining a new geological era, called {\it Anthropocene} \citep{crutzen2006} or {\it Capitalocene} \citep{malm2014,campagne2017}.
The global warming caused by these activities induces important changes in the climate system \citep{ipcc2021}.
 Working Group I of the IPCC, which assesses the physical science of climate change, summarises the latest advances in climate science to  understand the climate system and 
 assess climate change, by combining data from paleoclimate, observations and global circulation model (GCM) simulations. 
 The latter are based on  differential equations linked to the fundamental laws of physics, thermodynamics and chemistry. 
 GCMs simulate the evolution of various  climate variables on discretised tridimensional meshes with a typical horizontal resolution of 100 [km] or more.
The coupled model intercomparison project (CMIP) \citep{meehl2000, alexander2017} aims at comparing the performances of several dozen of GCMs  developed by different research centres, e.g.\ see Table  \ref{tab: CMIP_models} in Appendix.   
 As numerical experiments and approximations of the true climate system,  these GCMs can produce different climate responses to different given inputs, e.g.\  emission scenarios.
  To reduce model errors and gain robustness in signal detection,  GCMs are often analysed jointly.
 In particular, CMIP models have been used in the field of detection and attribution that aims at finding causual links    between the climate response and known external forcings 
 \cite[see, e.g.][]{Ribeseabc0671,Geert21,naveau20}.
As a yardstick,  the so-called  ``natural forcings" runs have not been    influenced  by human activities and were only driven by    external  forcings,
 {\it e.g.} solar variations, explosive volcanic eruptions  like Mont Pinatubo in 1991 
\cite[see, e.g.][]{Ammann10}. 
Such a numerical setup can be viewed as  a thought-experiment and it corresponds to a counterfactual world, but not to the observed one.
In contrast, a factual world is produced by integrating all forcings, including rising greenhouse gazes, and 
factual runs aims at reproducing the observed climatology over the last century. 
Future periods, say 2071--2100, can also be explored with GCMs  but 
future forcing  and emission scenarios  need to be chosen.
For example,  {\tt RCP8.5} for CMIP5 \citep{ipcc2013} and {\tt SSP5-8.5} for CMIP6 \citep{ipcc2021} will be analysed in this paper. 
 In this context, a natural question is to wonder how the climate system will change under these   scenarios.

Due to their large societal and economical impacts, a vast literature has be dedicated to answering this question for extreme events. 
In particular, heavy rainfall and heatwaves have received a particular attention, see chapters 10 and 11 in the Working Group I contribution of \cite{ipcc2021} report.  
In this paper, we focus on  annual maxima of daily  precipitation from 1850 to 2100 
 provided by the  factual (all forcings) and counterfactual (natural forcings only)  models 
 listed in Table \ref{tab: CMIP_models} of the Appendix. 
 Note that  our main  climatological goal is not to directly assess changes in  heavy rainfall intensities and frequencies, but rather to 
detect how spatial patterns   (clusters) of yearly maxima of daily precipitation could be modified by   anthropogenic forcing.

\begin{figure}[ht!]
    \centering
    \includegraphics[width = .9\textwidth]{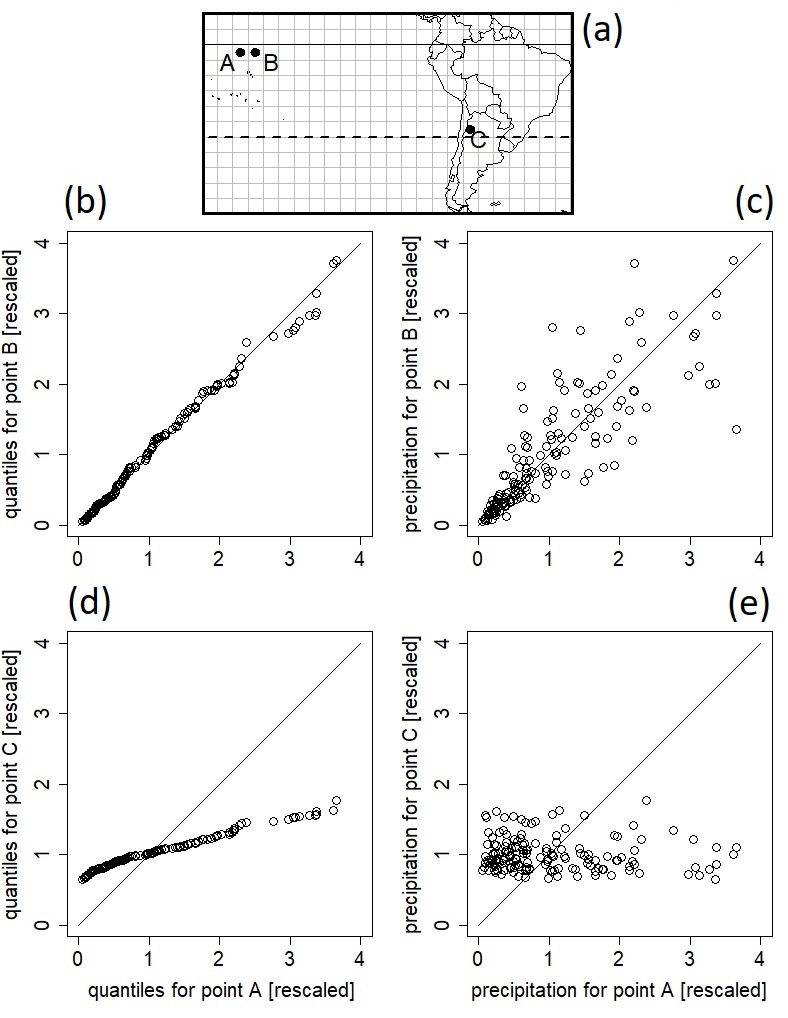}
    \caption{Localisation (a), QQ-plots (b) and (d) and scatter plots (c) and (e) of annual precipitation maxima at three grid points A,B and C in the counterfactual run of the {\tt CCSM4} model (1850--2005).  Panels (b) and (d) show the QQ-plots of rescaled precipitation for pairs (A,B) (b) and (A,C) (d). Panels (c) and (e) display the (rescaled) scatter plots for the same pairs.}
    \label{fig: QQplots_CCSM4_histnat}
\end{figure}

 To model  yearly block maxima, one   classical statistical  approach  is to impose a parametric  
  generalised extreme value (GEV) distributions \citep[see {\it e.g.}][]{Coles2001,davison2012}. 
 For example, each grid point of each individual  CMIP model  could  be fitted with  a spatial structure embedded  within the GEV parameters  \citep[see, e.g.][]{kharin2013}. 
However, the  computational cost can be high (more than 200 years of precipitation data at thousands of grid points for 16 models), especially 
if the spatial dependence is included. 
Another aspect is the ease of  interpretation. 
Well defined spatial patterns (clusters)  in extreme precipitation are very useful for climatologists who can  interpret them according to known  physical phenomena
\cite[e.g.,][]{Pfahl17,Tandon18,Dong21}. 
For example, the so-called 
  regional frequency analysis (RFA) has been frequently used in hydrology, see \cite{dalrymple1960,hos05}, but it has been rarely implemented in a D\&A context, 
  especially within the CMIP repository. 
  The main idea of RFA is to identity homogeneous regions 
 with identical distributional features, up to normalising constants. 
 More precisely
two positive absolutely continuous random variables (r.v.)  $Y_1$ and $Y_2$ are said to be homogeneous if there exists a positive constant $\lambda$ such that
$$
Y_2  \overset{\mathrm{d}}{=} \lambda  Y_1,
$$
where $\overset{\mathrm{d}}{=}$ denotes equality in distribution. This condition can be reformulated in terms of their cumulative distribution functions (cdf)  $F_i(x)= \mathbb{P}(Y_i\leq x)$ with $i\in\{1,2\}$ as 
\begin{equation}\label{eq: stochastic}
 F_2( \lambda x) = F_1(x).
\end{equation}
Hence,   two climate model grid points are said to  belong to the same homogeneous region if they satisfy \eqref{eq: stochastic}.
To visually understand this condition within the CMIP archive, three  grid points, say  A, B and C,  from the CCSM4 counterfactual run  are plotted 
in panel (a) of Figure \ref{fig: QQplots_CCSM4_histnat}.
In panel (b), 
 ranked annual precipitation maxima (rescaled  by the empirical mean)  of point A are compared 
to the ones from point B. Panel (d) provides the same information but between point A and point C. 
It appears that 
%
points A and B are likely to satisfy  \eqref{eq: stochastic} and, consequently, could belong to the same   homogeneous region.
In contrast, 
the rescaled distribution at point A is much more heavy-tailed than at point C. 
This is not surprising because A and B are nearby and C far away from them.
Still,  panels (b) and (d)  only rely on the marginal behaviours, and pairwise  dependence information and/or covariates could help   
finding of homogeneous regions. 

\begin{figure}
    \centering
    \includegraphics[width = \textwidth]{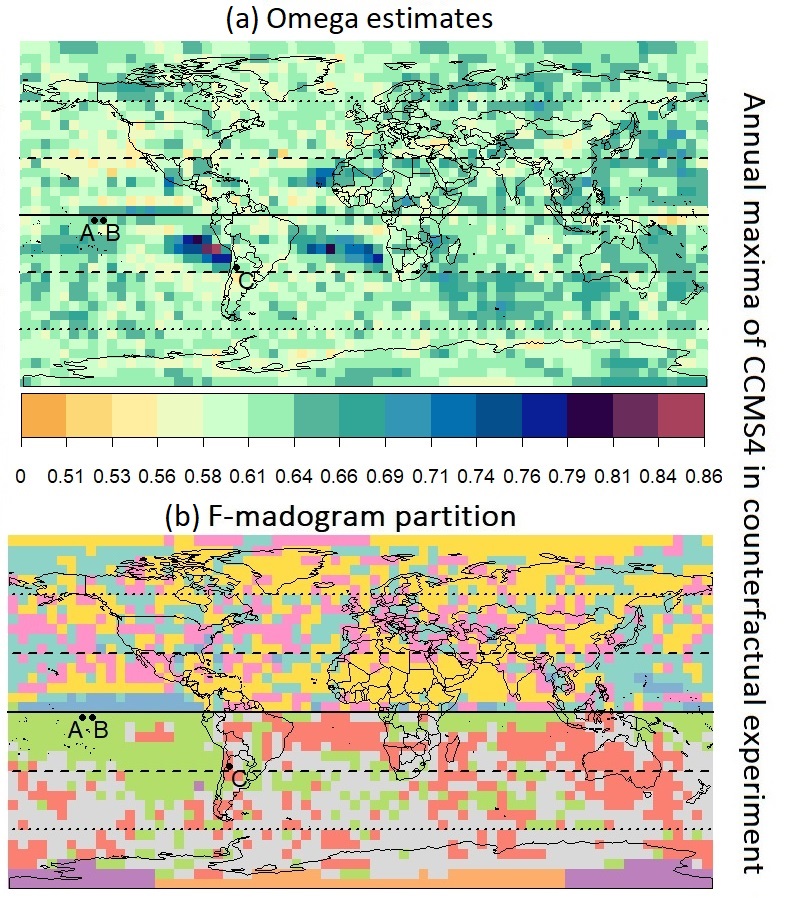}
    \caption{Two summaries of the structure of the precipitation annual maxima of counterfactual (1850--2005) {\tt CCSM4} model. (a) Pointwise  $\omega$ ratio \citep{legall21}. High values of $\hat{\omega}$ correspond to heavy tailed distributions. (b) Results of \texttt{PAM} clustering with the F-madogram distance \citep{bernard13}, with four clusters for each hemisphere separately. Each color corresponds to a cluster. 
  }
    \label{fig: GEV_Fmad_CCSM4}
\end{figure}

Various RFA techniques based on explanatory covariates \citep[e.g., see][for recent work]{asa18, fawad18} have been developed to identify homogeneous regions which rely on  station location features   and/or weather patterns to explain precipitation spatial distributions  \citep[see {\it e.g.}][]{ burn90,hos05, Evin16}.  
For example, \cite{toreti2016} let  scale parameters vary as a function of weather station locations. 
 However, selecting relevant covariates is constrained by their availability, expert subjectivity and the scale of the problem. 
 In particular, finding appropriate covariates for heavy rainfall patterns at the global scale is tedious. 
 In addition,  assessing  the homogeneity of  regions \citep{hos05} relies on specific moments like  skewness and kurtosis   
 that are not necessary robust (based on the spatial independence assumption).
 Other techniques bypass the use of covariates by only working with the data at hand, here precipitation  \citep{saf2009}. 
 For example, \cite{legall21} considered a ratio of probability weighted moments, see \cite{green79} and applied a clustering algorithm on this ratio.
 More precisely, 
this ratio, denoted $\omega \in [0,1]$,  is mean and scale invariant,  i.e. in compliance with \eqref{eq: stochastic}, and  
 it is a simple  increasing function of $\xi$  when rainfall extremes can be assumed to either follow a GEV or Pareto distribution  with shape parameter $\xi$. 
To illustrate the spatial variability of CMIP rainfall tail index (i.e. of $\omega$), panel (a) of Figure \ref{fig: GEV_Fmad_CCSM4}
displays the ratio $\omega$ at each grid point of a counterfactual {\tt CCSM4} annual maxima run.
Note that grid points A and B exhibit similar $\omega$ estimates, while grid point C differs (lighter tail).


All aforementioned RFA techniques has  one major drawback.
They rely on the  assumption of pairwise independence   or pairwise conditional independence (given the covariates). 
Note that  Eq.\eqref{eq: stochastic}  also constraints the marginal behaviour, but does not take into account of  any information about the spatial dependence strength. 
 Still, precipitation series at two nearby grid points  are likely to be dependent.
 To illustrate this point, we can go back to Figure \ref{fig: GEV_Fmad_CCSM4}. 
 Panels (b) and (e) display the scatter plots (rescaled by their means) between points A and B,  and between points A and C, respectively. 
 As expected from their local proximity, not only A and B have same similar marginals, but annual maxima of daily precipipation appears to 
be strongly  correlated. This information coupled with constraint \eqref{eq: stochastic} should  play an important role in improving RFA methods. 


Modelling the dependence structure in clustering algorithms   can be handled in different  ways depending   on the assumptions one is ready to make.
Fully non-parametric or parametric approaches   can be developed. Explanatory covariates can be included or difficult to find.  
For example, 
\cite{kim2019}  introduced a parametric approach based on copulas in the context of cluster detection in mobility networks. They grouped sites subject to intense traffic according to covariates ({\it e.g.} geographical), and checked the dependence strength within each cluster by fitting a multivariate Gumbel copula.
\cite{drees2019} and \cite{janssen2020} proposed approaches based on exceedances; after projecting observations onto the unit sphere, they reduced their dimension through $K$-means clustering \citep{janssen2020} and principal component analysis \citep{drees2019}.
 Finally, \cite{bernard13} applied a non-parametric approach based on the F-madogram to weekly precipitation maxima. 
The so-called F-madogram \citep{cooley2006} is defined by
\begin{equation}\label{eq: mado}
d= \frac{1}{2}  \mathbb{E} \left| F_1\left(Y_1\right) - F_2\left(  Y_2\right) \right|,
\end{equation}
where $Y_i$ is the continuous r.v.\ with cdf $F_i$.
It is a distance    which, by construction, is marginal-free because the r.v.\
$F_1\left(Y_1\right)$ and $F_2\left(Y_2\right)$ are both uniformly distributed on $[0,1]$. 
Note that if $Y_1$ and $Y_2$ are  equal in probability, the distance $d=0$. 
Whenever the bivariate vector $(Y_1, Y_2)$ follows a bivariate GEV distribution \citep[see {\it e.g.}][]{gumbel1960,tawn1988}, this distance can be interpreted as linear transformation of the extremal coefficient \citep[see {\it e.g.}][and Section \ref{sec: bi-GEV}]{cooley2006, naveau2009}. \cite{bernard13},  \cite{bador15} and later  \cite{saunders19} computed this distance to build a  
pairwise dissimilarity matrix that was  used as an input of a clustering algorithm.
In these two former studies, a partitioning around medoids (\texttt{PAM}) algorithm \citep{Kaufman1990} was applied whereas the latter used hierarchical clustering.
But, the RFA requirement defined by \eqref{eq: stochastic} was not imposed, and so the marginal differences between $Y_1$ and $Y_2$ 
were not taken into account. 
To visualise this issue within the  CMIP repository, it is simple to  cluster a counterfactual {\tt CCSM4} annual maxima run  with the \texttt{PAM}  
algorithm\footnote{In all our  CMIP analysis, \texttt{PAM}  was applied separately  to  the southern and northern  hemispheres. 
Global analysis (available upon request) were also made, but the climatological interpretation was not as clear as with the hemispheric scale.
Also, different numbers of clusters were investigated and basic criteria like the silhouette coefficient were computed. 
No particular number could be clearly identified. But, in terms of interpretation, 
four clusters appear as a reasonable compromise between climate understanding, visual simplicity and statistical criteria.} based on the distance $d$.  
The resulting map displayed in panel (b) of Figure \ref{fig: GEV_Fmad_CCSM4} shows 
a few spatially coherent structures, but,  overall is very patchy.   
In addition, panel (a) related to the marginals behaviour appears to be unrelated to panel (b)   that describes the spatial dependence.
This was expected from the F-madogram distance, but it would  make sense to cluster grid points that are both correlated but also the same type of marginal, 
see \eqref{eq: stochastic}, the essence of the RFA. 

To reach this goal, we propose the following work plan. 
In Section \ref{sec: RFA-mado}, we integrate the homogeneity condition \eqref{eq: stochastic}  into a new  definition of the F-madogram distance.
The properties of this new  dissimilarity, which we call RFA-madogram, is explained by analysing  a special case: the logistic  bivariate GEV model
in Section \ref{sec: bi-GEV}.
A non-parametic estimator of the RFA-madogram is proposed  and its asymptotic consistency in law  is detailed  in 
Section \ref{sec: inference}.
Concerning the CMIP database, we compute, in Section \ref{sec: appli_RFAmad},  a  RFA-madogram dissimilarity matrix on  
annual maxima of daily precipitation  for  each  CMIP models listed  in Table 
\ref{tab: CMIP_models}, and then cluster them with the \texttt{PAM} algorithm. 
 Finally, we propose a method to build a ``central" partition that summarises the partitions obtained for each model and compare the spatial patterns obtained for counterfactual (1850--2005) and factual (2071--2100) experiments.
Section  \ref{sec: conclusion} concludes  the paper by  providing a short discussion.


\section{Joint modelling of dependence and homogeneity}\label{sec: RFA-mado}

\subsection{RFA-madogram}

To introduce homogeneity criteria, see Eq.\eqref{eq: stochastic}, into distance defined in Eq.\eqref{eq: mado}, we propose to define and study the following expectation 

\begin{equation}\label{eq: mado-RFA}
 D(c,Y_1, Y_2)= \frac{1}{2} \mathbb{E} \left| F_2\left(c  Y_1 \right) - F_1\left(  \dfrac{Y_2}{c}  \right) \right|,
\end{equation}
 where $c > 0$ is a normalising positive constant.
The $D(c,Y_1,Y_2)$ is always non-negative
and equal to zero for $c= \lambda$ when $Y_2 \overset{a.s}{=} \lambda Y_1$. The homogeneous regions are not defined {\it a priori}, so the existence of $\lambda$ and its value are not known.
We denote 
$$
c_{12}^* = \mbox{argmin}\{ D(c,Y_1,Y_2): c >0 \}.
$$
Note that $D(c,Y_{1},Y_{2}) = D\left(\dfrac{1}{c}, Y_{2}, Y_{1}\right)$, for all positive $c$. Therefore, $c_{12}^* = \dfrac{1}{c_{21}^*}$ .
 The particular case of equality in distribution, $Y_1 \overset{d}{=} Y_2$, corresponds to the case where $c_{12}^* = c_{21}^* = 1$.
An important feature of Eq.\eqref{eq: mado-RFA} is that, under the homogeneity condition of Eq.\eqref{eq: stochastic},  
$$
  D(\lambda,Y_1,Y_2) = d(Y_1,Y_2),$$
where $d$ is the classical F-madogram, see Eq.\eqref{eq: mado}.
To simplify notations, $D$ or $D(c)$ will be a shortcut for  $D(c,Y_1,Y_2)$.

The key point from a RFA point of view is  that, if Eq.\eqref{eq: stochastic} is satisfied, $D$ behaves as the classical F-madogram distance. 
Note that  $D$ is not a true distance, but a dissimilarity. The triangle inequality is satisfied under homogeneity condition but may not be valid in general. 
Still, $D$ captures information about the extremal dependence like the F-madogram, and, in addition, it encapsulates    marginal information concerning the departure from Eq.\eqref{eq: stochastic}. 
More precisely, one can show (see Appendix A for the proof) that 
\begin{equation}\label{eq: dist diff}
2 \left|  d-D \right|  \leq  \mathbb{E}\left[\Delta(c,Y_1)\right] +   \mathbb{E}\left[\Delta(c,Y_2/c)\right],
\end{equation}
where the  function  
$
\Delta(c,x) =\left|  F_2\left( c x \right)    - F_1\left(x \right)   \right|$
measures the difference  between the rescaled cdfs.

To deepen our understanding of $D$, we comment on the special case of a bivariate-GEV distributions. 

\subsection{RFA-madogram for bivariate GEVs}\label{sec: bi-GEV}
In this section, we suppose that the bivariate vector $(Y_1,Y_2)$ follows a  max-stable distribution \citep{Coles2001, fougeres2004, guillou2014} with dependence function $V(.,.)$
$$
    \PP (Y_1 \leq x; Y_2 \leq y) = \exp \left[-V\left\{\dfrac{-1}{\log F_1 (x)}, \dfrac{-1}{\log F_2 (y)}\right\} \right],
$$
where $F_i$ corresponds to a GEV marginal cdf.
If $F_i(x) = \exp \left\{ - \left( \dfrac{x}{\sigma _i}\right)^{-1/\xi_i} \right\}$ 
 with $\xi_1 = \xi_2 =\xi$, then the  equality $Y_2 \overset{d}{=} \dfrac{\sigma_2}{\sigma_1} Y_1$ holds and 
 we are in the homogeneity case.
 The shape parameter $\xi$ describes the common upper-tail behaviour. The larger $\xi$ is, the heavier the upper-tail of the distribution.
Although complex, Eq.\ \eqref{eq: D bi-GEV} in Appendix \ref{sec: app D(c)}, summarises how  $D(c)$ 
can be expressed   in function of $V(.,.)$ and the marginal parameters. 

\begin{figure}
    \centering
    \includegraphics[width= \textwidth]{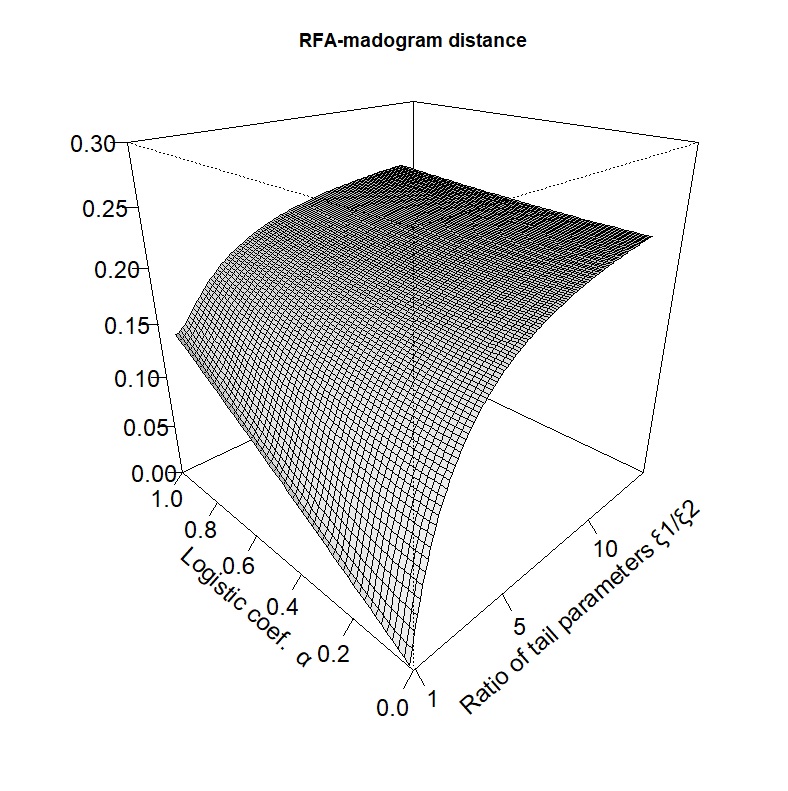}
    \caption{Distance ($z$-axis) $D$ defined in Eq.\eqref{eq: mado-RFA} in the logistic bivariate GEV model example. The normalising coefficient is chosen as the optimal one, $c^*$. The $x$ and $y$-axis indicate the dependency coefficient $\alpha$ in the logistic dependence, see Eq.\eqref{eq: log_dep} and the ratio of tail parameters {\it i.e.} the homogeneity of the two r.v. A ratio equal to one corresponds to the homogeneous case. A ratio equal to 10 can be illustrated by the realistic case of $\xi_1 = 0.1, \xi_2 = 0.01$.}
    \label{fig: RFAmad_biGEV}
\end{figure}

 To simplify  the dependence strength interpretation, it is common to focus on the extremal coefficient defined as the scalar $\theta_{12}$  such that 
$$
    \PP (Y_1 \leq u, Y_2 \leq u) = \left\{\PP(Y_1 \leq u) \PP(Y_2 \leq u)\right\}^{\frac{\theta_{12}}{2}}.
$$
It is equal to  $\theta_{12} = V(1,1)$. 
If $Y_1$ and $Y_2$ are independent, then $\theta_{12}=2$, while if they are fully dependent, then $\theta_{12}=1$.
Appendix \ref{sec: append homog} provides the mathematical details to link the extremal coefficient with $D(c)$. 
It allows to find an optimal value for rescaling parameter $c_{12}^*$. 
For example, it is possible to show  that 
$
    c_{12}^* = \dfrac{\sigma_2}{\sigma_1} = \lambda.
$
for  the logistic GEV model,
\begin{equation}\label{eq: log_dep}
    V(x,y) = \left(x^{-\frac{1}{\alpha}}+y^{-\frac{1}{\alpha}}\right)^\alpha, \mbox{ with } \alpha >0.
\end{equation}
In particular, 
the value of the dissimilarity $D(c_{12}^*)$ 
can be plotted as a function of the logistic coefficient $\alpha$ and of the ratio ${\xi_1}{/\xi_2}$. 
From Figure \ref{fig: RFAmad_biGEV}, one can see that the full dependence case corresponds to $\alpha \approx 0$, and the independence case to $\alpha = 1$.
In addition, the ratio ${\xi_1}/{\xi_2}$ varies between 1 (homogeneity case) and 10, i.e. cases with  $\xi_1 =0.1$ and $\xi_2 = 0.01$.
The dissimilarity is  small when both the dependence is strong and the marginals are homogeneous (leftmost corner). 
Large dissimilarities correspond to the opposite cases, a near independence and/or strong heterogeneity in the shape parameters (rightmost corner). 
Note also that, 
as the homogeneity and the dependence strength decrease jointly, 
dissimilarity  increases (concavity of the surface). 
These features correspond to our goal that, given the same dependence strength,  the price to pay  is high  when the RFA condition \eqref{eq: stochastic} does not hold. 
In other words, our aim  to cluster grid points that are jointly strongly dependent and in compliance with \eqref{eq: stochastic}
seems, at least conceptually,  to have been reached. The remaining question is to know if this strategy works  in practice with the CMIP archive. 
To answer this, we need to first check that a non-parametric estimator can be developed and its asymptotic properties can be  well understood. 

\section{RFA-madogram Inference}\label{sec: inference}
Given $\spX\subset \real^n$ and  $n\in \nat$, let 
$\ell^{\infty}(\spX)$ denote the spaces of 
bounded real-valued functions on $\spX$.
For $f\!\!\!:\spX \rightarrow \real$, let $\|f\|_{\infty}=\sup_{\bx \in \spX} |f(\bx)|$. 
The arrows ``$\conAS$'', ``$\conD$'', and ``$\conW$''  denote almost sure convergence, convergence in distribution of random vectors
\citep[see][ Ch.\ 2]{vaart98} 
and weak convergence of functions in $\ell^{\infty}(\spX)$ 
\citep[see][ Ch.\ 18--19]{vaart98}, respectively. 
Let $L^2(\spX)$ denote the Hilbert space of square-integrable functions $f\!\!\!: \spX \to \real$,
with $\spX$ equipped with $n$-dimensional Lebesgue measure; the $L^2$-norm is denoted by 
$\|f\|_2=\left\{\int_{\spX} f^2(\bx) \, \diff\bx\right\}^{1/2}$. 

In this section, given a  sample of bivariate observations, say $\left( {\bf Y}_1, \dots, {\bf Y}_n \right)^t$, we focus on  the asymptotic properties of two RFA-madogram estimators. Two cases can be studied:  when the marginal distributions, $F_1$ and $F_2$, are known or when we need to use their empirical estimator, say $\widehat{F}_1$ and $\widehat{F}_2$.  
In both cases, the copula function of the bivariate vector $(Y_1,Y_2)^t$, say $C(u_1,u_2)$,   that captures the dependence structure needs to be inferred. 
To derive our asymptotic results, we adapt the main ingredients of Theorem 2.4 from \cite{marcon2016}  to our settings, see Appendix B for details.   
With the notation  
$$
a_c(u) = F_2\left\{c F^{\leftarrow}_1(u)\right\}, 
$$
we can write 
$$
D(c)= \frac{1}{2} \mathbb{E} \left| a_c\left(U_1\right) - a^{\leftarrow }_c\left( U_2  \right) \right|,
$$
where the bivariate vector ${\bf U}=(U_1,U_2)^t$ follows  
 the copula $C({\bf u})$. 
This leads us to the estimators 
%
$$
D_n(c)=\frac{1}{n} \sum_{i=1}^n D_c\left({\bf U}_i\right)
\mbox{, with } 
{\bf U}_i = (F_1(Y_{1,i}),F_2(Y_{2,i}))^t \mbox{ and }
D_c\left({\bf U}_i\right) = \left| a_c\left(U_{1,i}\right) - a^{\leftarrow }_c\left( U_{2,i} \right) \right|.
$$
If $F_1$ and $F_2$ are unknown and are replaced by their empirical estimators, we have, with 
$
\hat{a}_c(u) = \hat{F}_2\left\{c \widehat{F}^{\leftarrow}_1(u)\right\}, 
$ 
$$
\widehat{D}_n(c)=\frac{1}{n} \sum_{i=1}^n \hat{D}_c\left(\widehat{{\bf U}}_i\right)
\mbox{, with } 
\widehat{{\bf U}}_i = \left\{\widehat{F}_1(Y_{1,i}),\widehat{F}_2(Y_{2,i})\right\}^t  
\mbox{ and  } 
\widehat{D}_c\left(\widehat{{\bf U}}_i\right) = \left| \hat{a}_c\left(\widehat{U}_{1,i}\right) - \hat{a}^{\leftarrow }_c\left( \widehat{U}_{2,i} \right) \right|.
$$
In practice, $\widehat{D}_n(c)$ is directly computed from the expression 
$$
\widehat{D}_n(c)= \frac{1}{n} \sum_{i=1}^n \left| \widehat{F}_2( c Y_{1,i}) - \widehat{F}_1(Y_{2,i }/c) \right|.
$$
Still, the  definition  of $\widehat{D}_n(c)$ with $\widehat{{\bf U}}_i$ 
facilitates the derivation of theoretical results by leveraging 
existing properties of the empirical copula 
$$
C_n({\bf u}) =  \frac{1}{n} \sum_{i=1}^n   {\mathbb{I}}({\bf U}_i\leq {\bf u}) 
\mbox{ and by  writing } 
D_n(c)= \int_{[0,1]^2} D_c\left({\bf U}\right) dC_n\left({\bf u} \right). 
$$
In particular,  the following   classical  smoothness condition on   copula $C$ is needed,  
see Example~5.3 in \cite{seger12} for details. \\
{\bf Condition (S)}\\
For every $i \in \{1,2\}$, the partial derivative of $C$ with respect to $u_i$ exists and is continuous on the set $\{\bu\in[0,1]^2: 0< u_i<1\}$.

%
%
\begin{proposition}\label{prop:prop_multimado}
Let $\left( {\bf Y}_1, \dots, {\bf Y}_n \right)^t$ be $n$  independent and identically distributed random vectors whose common distribution has continuous margins and  a  copula function $C$ that satisfies condition (S).

Let $\CB$ be a $C$-Brownian bridge, that is, a zero-mean Gaussian process on $[0,1]^2$ with continuous sample paths and with covariance function given by
\begin{equation}\label{eq:covariance}
\mathbb{C}ov(\CB(\bu),\CB(\bv))=C(\bu\wedge\bv)-C(\bu) \, C(\bv),\qquad \bu,\bv\in[0,1]^2.
\end{equation}
Here $\bu \wedge\bv$ denotes the vector of componentwise minima. We define the Gaussian process $\copula$ on $[0, 1]^2$ by
\begin{equation*}\label{eq:cop_proc}
\copula(\bu)=\CB(\bu)- \frac{\partial C}{\partial u_1}\, \CB(u_1,1)
- \frac{\partial C}{\partial u_2}\, \CB(1,u_2)
\end{equation*}
%
Then we can write that 
\begin{itemize}
\item[a)]
We have $\norm{D_n(c) - D(c) }_\infty \to 0$ almost surely as $n \to \infty$. Moreover, 
as $n \to \infty$,
\begin{multline*}
\sqrt{n}\left\{D_n(c)-D(c)\right\}\conW \\
\frac{\left\{1+D(c)\right\}^2}{2}
\left[ \int_0^1 \left\{\CB(a^{\leftarrow}_c(x),1)- \CB(a^{\leftarrow}_c(x),a_c(x))\right\} dx + 
\int_0^1 \left\{\CB(1, a_c(x)) - \CB(a^{\leftarrow}_c(x),a_c(x))\right\}dx 
\right]
\end{multline*}
\item[b)]
We have $\norm{\widehat{D}_n(c) - D(c) }_\infty \to 0$ almost surely as $n \to \infty$, and 
as $n \to \infty$,
\begin{equation*}\label{eq:wc_pick}
\sqrt{n}\left\{\widehat{D}_n(c)-D(c)\right\}\conW 
\left[-\left\{ 1+D(c)\right\}^2\int_0^1 \copula\left\{a^{\leftarrow}_c(x),a_c(x)\right\} \, 
\diff x\right]_{c >0}.
\end{equation*}
%
\end{itemize}
\end{proposition}
%

%
%
%


\section{Analysis of CMIP precipitation for 16 models under two experiments }\label{sec: appli_RFAmad}
We now apply the RFA-madogram to the problem of partitioning annual precipitation maxima from 16 CMIP GCMs (see Table \ref{tab: CMIP_models} in Appendix) into homogeneous regions. 
For each hemisphere of a given GCM run,  we estimate  the dissimilarity matrix $D(c^*)$ (Eq.\eqref{eq: mado-RFA}) between each pair of grid points.
To cluster from a dissimilarity matrix, the  \texttt{PAM} clustering algorithm is implemented as it is fast, adapted to  max-stable distributions  
\citep{bernard13}, and 
it does not require  the triangle inequality  \citep{schubert2021}. 
The counterfactual (1850--2005) and factual (2071--2100) runs are analysed separately and later compared to identify possible differences.\\
With 16 partitions in four clusters for each 16 counterfactual (factual) hemispheric runs,  GCM in-between-model error becomes an issue in terms of interpretation. 
We therefore summarise them in one ``central" partitions, which we obtain in two steps. 
First, partitions for each counterfactual hemispheric runs  are relabelled so as to minimise the pairwise difference between two partitions by taking each (alternatively) as reference score. As an example with five grid points, the partitions \{1 1 1 2 2 3\} and \{3 3 3 1 1 2\} are equal up to the permutation $(1, 3, 2)$. Then, we compute the probability of each grid point to belong to each of the clusters, and associate the corresponding grid point to the cluster of highest probability. For instance, grid point B is assigned to cluster 1 for 6 models out of 16, to cluster 2 for 9 models and to cluster 3 for only one model. In the so-called central partition, B is then assigned to cluster 2 with probability 9/16. Partitions for the factual experiment are relabelled in order to minimise the difference with the counterfactual central partition.

\begin{figure}
    \centering
    \includegraphics[width=\textwidth]{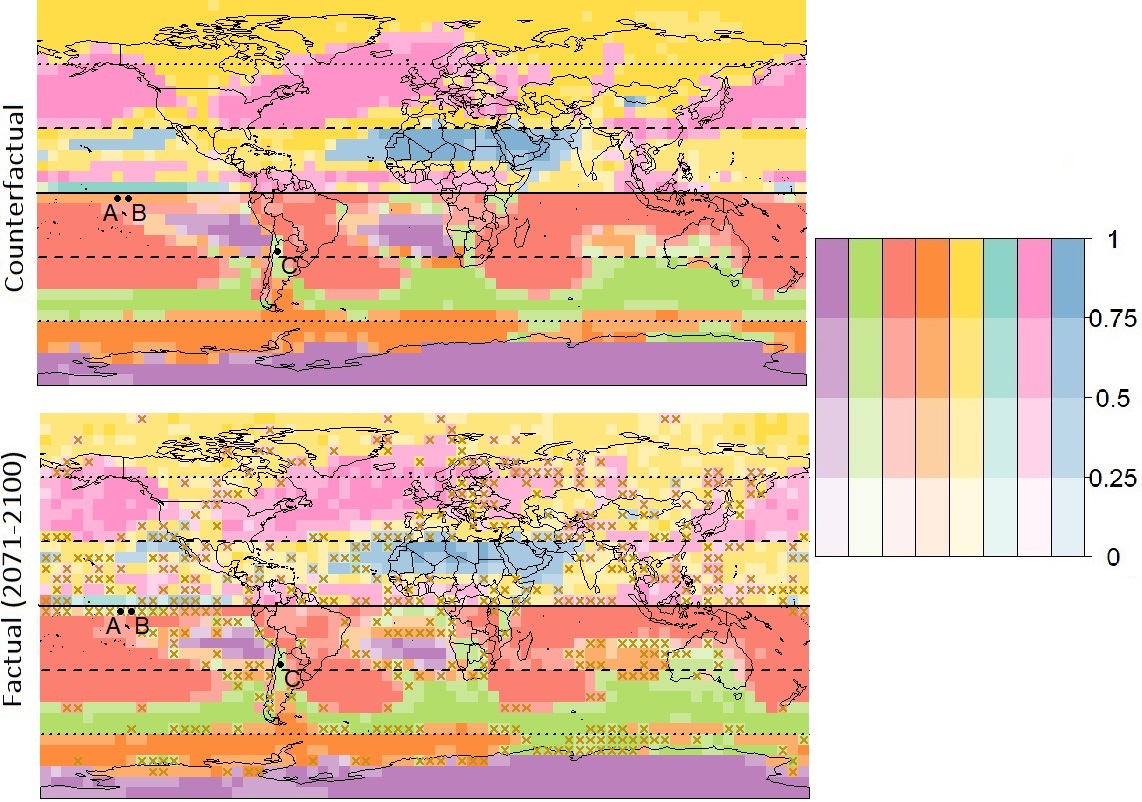}

    \caption{Central partitions of CMIP models (with four clusters for each hemisphere), for (top) the counterfactual experiment (1850--2005) and (bottom) the factual experiment (2071--2100). Each colour corresponds to a cluster, with the shade indicating the probability of belonging to that cluster. In the bottom map, brown crosses indicate points where the most likely cluster is different between the counterfactual and the factual experiments.}
    \label{fig: CMIP_NS_RFAmad8}
\end{figure}

For example,  Figure \ref{fig: CMIP_NS_RFAmad8} shows  the central partitions in  four clusters by  hemisphere. 
Intense colours correspond to points that belong to the same cluster in most, if not all, model partitions. Beginning with the counterfactual experiment, we first note that the clusters are very coherent spatially, in stark contrast to marginal- ($\omega$) and dependence-based (F-madogram) partitions (Figure \ref{fig: GEV_Fmad_CCSM4}), even though no geographical covariates were used in the clustering.\\
The Northern Hemisphere is dominated by two clusters (pink and yellow), with two others (blue and turquoise) with limited spatial extent. The distribution is more even in the Southern Hemisphere, and also more zonally symmetric.\\
These partitions, driven both by homogeneity and dependence, are generally consistent with precipitation climatology. In the Northern Hemisphere, the pink cluster extends over the storm track regions of the North Atlantic and Pacific Oceans, and over the Inter-Tropical Convergence Zone (ITCZ) around $10^\circ$N. The blue cluster covers the dry subtropics above the Sahara, Southwest Asia and southwest of North America. The turquoise cluster is located in the dry zone above the cold Pacific tongue, while the yellow cluster includes most regions with semi-arid and continental climates. Still, it also includes monsoon-dominated regions ({\it e.g.}, India) and the dry Arctic.\\
In the Southern Hemisphere, arid regions in Antarctica and in the dry descent regions at the eastern edge of the subtropical anticyclones are grouped together in the purple cluster, while the red cluster covers much of the wet tropics. The orange and green clusters correspond to the Southern Hemisphere storm track.

\begin{figure}[!ht]
   \centering
  \includegraphics[width = \textwidth]{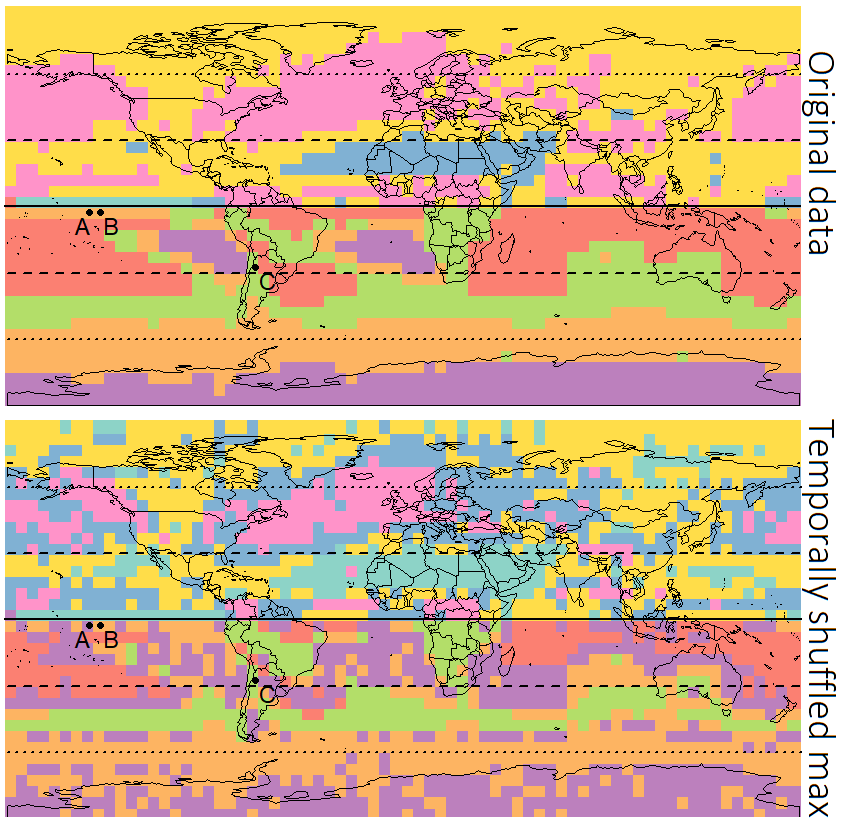}
   \caption{Partition of {\tt CCSM4} model in the counterfactual experiment based on the RFA-madogram dissimilarity $D(c^*)$ and \texttt{PAM} algorithm, for (top) original data, and (bottom) data randomly shuffled in time at each grid point. The clustering algorithm is applied to each hemisphere independently.}
    \label{fig: CCSM4_shuffle}
\end{figure}

Most of the clusters appear to be quite robust across GCMs. Notable exceptions are the ITCZ regions in the Northern Hemisphere, and the equatorial Pacific and the eastern Indian Ocean west of Australia in the Southern Hemisphere. This lack of robustness may be due to the choice of cluster number. In any case, some differences are expected across GCMs, as they differ in their representation of storm tracks, monsoons or ITCZ location and dynamics.\\
At first order, it appears that homogeneity of the distributions plays the dominant role, with arid or wet regions grouped together in both hemispheres. Still, the clustering is by design not only based on marginal distributions but also on dependence strength. To measure the importance of dependence in the spatial structure, we apply our clustering algorithm to temporally shuffled annual maxima at each grid point. This removes any spatial dependence between variables while preserving their marginal distributions.\\
The results of Figure \ref{fig: CCSM4_shuffle} for the {\tt CCSM4} model show a much less spatially coherent partition for the shuffled data. The dependence thus plays an important role in the coherence of the partition. This role can be further quantified by computing the relative difference between RFA-madogram on shuffled and non-shuffled data (with respect to the medoids). For about $2/3$ of the grid points, the RFA-madogram takes lower values on the non-shuffled data, in particular near the medoids.

We now turn to the comparison of the central partitions between the factual and counterfactual experiments. The overall partition structure is very similar in both experiments (Figure \ref{fig: CMIP_NS_RFAmad8}). The clusters are better defined in the counterfactual experiment (\textit{i.e.} cluster probabilities closer to 1) because the sample size is much larger than for the factual experiment (155 \emph{versus} 30 years). Globally, differences between the two central partitions are not significant compared to variability of model partitions compared to the central partition for either the factual or the counterfactual experiment (not shown). Hence, we cannot conclude to more spatial pattern variability in the factual world.\\
The most likely cluster changes for a number of grid points, however, as indicated by crosses on the bottom panel of Figure \ref{fig: CMIP_NS_RFAmad8}. In the Northern Hemisphere, the pink (humid) and blue (arid) clusters expand slightly Northwards. More specifically, the probability of a given grid point to belong to the pink cluster generally increases at high latitudes, while the probability to belong to the blue cluster increases around the 25$^{\circ}$N latitude. In the Southern Hemisphere the green cluster (humid) also expands Southwards.\\
While the resolution of our analysis is rather low (5$^{\circ}$), these differences are consistent with the expected polewards shift of major climate zones under climate change, particulary the arid subtropics and the storm track regions of both hemispheres \citep{Scheff12}.

\section{Conclusion}\label{sec: conclusion}

When considering multivariate data, extreme value theory can be difficult to handle. Reducing the dimensionality of extreme precipitation data set is then a challenging task.
Our main goal in this work was to show that a simple and fast clustering approach based on an interpretable dissimilarity could highlight climatologically coherent regions.

The proposed approach coupled the main RFA idea, {\it i.e.}\ a normalising factor, with the dependence structure via the F-madogram.
The introduced  dissimilarity   has links with extreme value theory via the extremal coefficient and tail parameters. The RFA-madogram neither  requires estimating any marginal parameters nor  dependence parameters. 
It is fully data-driven and bypasses the need of selecting relevant covariates or dependence structure. 

Our analysis of annual maxima of daily precipitation from each CMIP model provides more spatially coherent hemispheric regions than some other  non-parametric methods focusing on only one aspect (either homogeneity or dependence). 
Another contribution of this work is the handling of multi-partitions as  our selected CMIP set  has 16 GCM runs. 
Our combining approach  enables   us  to compare 
one   multi-model partition of the factual (all forcings) world with 
another multi-model partition of  counterfactual (natural forcings) world.
It appears that spatial variability between all models for the factual (resp. counterfactual) experiment appears to be significantly higher than between the two factual and counterfactual experiments. 


In this work, we focus on the spatial structure of annual maxima precipitation in CMIP models, and on the  forcing impact. We did not directly study the changes in rainfall distributions and frequencies. 
One interesting perspective would be to model  precipitation intensities  and dependence structure within each cluster.
This could be useful for the D\&A community. 
Another aspect is that the statistical approach developed therein is  easy-to-implement and flexible, e.g. it can be used on non-gridded products. 
For example, it   could be applied to    large  weather networks, reanalysis (ERA 5) and radar  products.
Such datasets have finer spatial resolution scales than GCMs, and the dependence structure could  be stronger, and consequently 
the analysis of heavy rainfall spatial patterns at fine spatial scales  improved.

\newpage

\section*{Acknowledgement}
Within the CDP-Trajectories framework, this work is supported by the French National Research Agency in the framework of the ``Investissements d’avenir” program (ANR-15-IDEX-02). 

Part of this  work was supported by the  DAMOCLES-COST-ACTION on compound events, the French national program (FRAISE-LEFE/INSU
 and  80 PRIME CNRS-INSU), and the European H2020 XAIDA (Grant agreement ID: 101003469). 
 The authors also acknowledge the support of the French Agence Nationale de la Recherche (ANR) under reference ANR-20-CE40-0025-01 (T-REX project), 
 and the ANR-Melody.

 
\begin{appendix}


\begin{table}
    \caption{\label{tab: CMIP_models} List of 16 CMIP models considered, with institutions, belonging countries and native horizontal resolution (longitude by latitude in degree). AOR (UoT): Atmosphere and Ocean Research Institute (The University of Tokyo);  CSIRO: Commonwealth Scientific and Industrial Research Organisation; DOE: Department of Energy; JAMSTEC: Japan Agency for Marine-Earth Science and Technology; NIES: National Institute for Environmental Studies; NSF: National Science Foundation. Most models come from the CMIP phase 5, those coming from phase 6 are indicated by $^*$. In this paper, models are regridded to a resolution of $5 ^\circ$ x $5^\circ$. }
    
    \fbox{%
    \begin{tabular}{|cccc|}
    \hline
    &&&\\
       \large{Model}  &  \large{Institute} & \large{Country} & Resolution \\
       &&&\\
       \hline  
        &&&\\

       {\tt CanESM2} & Canadian Centre for Climate Modelling and Analysis & Canada & $2.8^{\circ}$ x $2.8^{\circ}$\\ 
        {\tt CanESM5$^*$} & ~&~ & $2.8^{\circ}$ x $2.8^{\circ}$  \\
        &&&\\
     
        {\tt CCSM4} & National Center for Atmospheric
Research (NCAR) & USA & $1.3^{\circ}$ x $0.9^{\circ}$\\ 
&&&\\
        {\tt CESM1-CAM5} & NSF, DOE and NCAR & USA & $1.3^{\circ}$ x $0.9^{\circ}$ \\
        &&&\\
        {\tt CNRM-CM5} & Centre National de Recherches M\'et\'eorologiques & France & $1.4^{\circ}$ x $1.4^{\circ}$\\
        {\tt CNRM-CM6-1$^*$} & ~ & ~& $1.4^{\circ}$ x $1.4^{\circ}$ \\
        &&&\\
     {\tt ACCESS1-3}  & CSIRO and Bureau of Meteorology & Australia & $1.9^{\circ}$ x $1.3^{\circ}$ \\
           {\tt CSIRO-Mk3-6-0}  & ~ & ~ & $1.9^{\circ}$ x $1.9^{\circ}$ \\
    
        &&& \\
        {\tt IPSL-CM5A-LR} & Institut Pierre Simon Laplace & France & $3.8^{\circ}$ x $1.9^{\circ}$\\
        {\tt IPSL-CM5A-MR} & ~ & ~ & $2.5^{\circ}$ x $1.3^{\circ}$\\ 
        {\tt IPSL-CM6A-LR$^*$} & ~ & ~ & $2.5^{\circ}$ x $1.3^{\circ}$\\
        &&&\\
        {\tt MIROC-ESM} & JAMSTEC, AOR (UoT), NIES & Japan & $2.8^{\circ}$ x $2.8^{\circ}$\\
        {\tt MIROC-ESM-CHEM}  & ~& ~ & $ 2.8^{\circ}$ x $ 2.8^{\circ} $ \\
        &&&\\
        {\tt MRI-CGCM3}  &  Meteorological Research Institute & Japan & $1.1^{\circ}$ x $1.1^{\circ}$ \\
        {\tt MRI-ESM2-0$^*$} & ~ & ~ & $1.1^{\circ}$ x  $1.1^{\circ}$ \\
        &&&\\
        {\tt NorESM1-M} & Norwegian Climate Centre & Norway & $2.5^{\circ}$ x $1.9^{\circ}$ \\
         &&&\\
        \hline
        
    \end{tabular}}

\end{table}

\section{Proof of Eq. \eqref{eq: dist diff}} 
We can write that 
\begin{eqnarray*}
2 D(c) &=&\mathbb{E} 
\left| F_2\left( c Y_1\right) - 
F_1\left(Y_1\right) 
+ F_1\left(Y_1\right) - F_2\left(  Y_2\right) 
+  F_2\left(  Y_2\right) -F_1\left(Y_2/c \right) \right| \\
&\leq& \mathbb{E} \left| F_2\left( c Y_1\right) -  F_1\left(Y_1\right)   \right| 
+ \mathbb{E} \left| F_1\left(Y_1\right) - F_2\left(  Y_2\right)  \right|  
+  \mathbb{E} \left| F_2\left(  Y_2\right) -F_1\left(Y_2/c \right)  \right| ,\\
&\leq&   2 d+  \mathbb{E}\left[\Delta({c},Y_1) \right] +   \mathbb{E}\left[\Delta({c},Y_2/c) \right]
\end{eqnarray*}
In the same way, we can write that 
\begin{eqnarray*}
2 d&=&
\mathbb{E} \left| F_1\left(Y_1\right) -  F_2\left( c Y_1\right) + F_2\left( c Y_1\right)  - F_1\left(Y_2/c \right) 
+F_1\left(Y_2/c \right) - F_2\left(  Y_2\right)  \right|,  \\
&\leq & 2 D(c)  + \mathbb{E}\left[\Delta({c},Y_1)\right] +   \mathbb{E}\left[\Delta({c},Y_2/c)\right].
\end{eqnarray*}
It follows that the inequality expressed in Eq.\eqref{eq: dist diff} is valid. \hfill $\square$

\section{Proof of Proposition \ref{prop:prop_multimado}} 
Let $a(u)$ be any continuous non-decreasing function from $[0,1]$ to $[0,1]$ and denote its inverse by $a^{\leftarrow}(u)$. 
The map
\begin{equation*}
\phi : \ell^{\infty}([0,1]^2) \to \ell^{\infty}([0,1]) : f \mapsto \phi(f)
\end{equation*}
defined by
$$
(\phi(f))(a) = \frac{1}{2} \left( \int_0^1 f\left(a^{\leftarrow}(u),1 \right) du + 
   \int_0^1 f\left(1,a(u) \right) du \right) - \int_0^1 f\left(a^{\leftarrow}(u),a(u) \right) du
   $$
is linear and bounded, and therefore continuous. 
To continue, we need the following lemma.

\begin{lemma}\label{lemma: H}
For any cumulative distribution function $H$ on $[0,1]^2$ and for any non-decreasing function $a(.)$ on $[0,1]$, the function
$$
\delta \left({\bf u} \right) = 
 \frac{1}{2}  \left| a(u_1)  - a^{\leftarrow}(u_2)  \right|
$$
satisfies 
\begin{equation}\label{eq:lemma}
   \int_{[0,1]^2}  \delta\left({\bf u} \right) dH \left({\bf u} \right)= (\phi(H))(a).
\end{equation}
\end{lemma}
Proof of Lemma \ref{lemma: H}:
Note that 
$$
\delta \left({\bf u} \right) = \max \left( a(u_1), a^{\leftarrow}(u_2) \right)
-\frac{1}{2}\left( a(u_1) +  a^{\leftarrow}(u_2) \right).
$$
For any ${\bf u} \in [0,1]^2$, we have 
$$
\max \left( a(u_1), a^{\leftarrow}(u_2) \right) =
1- \int_0^1 {\cal I}\left(u_1 \leq a^{\leftarrow}(u), u_2 \leq a(u) \right) du 
$$ 
and 
$$
\frac{1}{2}\left( a(u_1) +  a^{\leftarrow}(u_2) \right)
= 1-\frac{1}{2}\left( 
    \int_0^1 {\cal I}\left(u_1 \leq a^{\leftarrow}(u) \right)du
    +
    \int_0^1 {\cal I}\left(u_2 \leq a(u) \right)du.
    \right)  
$$
Substracting both expressions and integrating over $H$ implies 
\begin{eqnarray*}
\int_{[0,1]^2}  \delta\left({\bf u} \right) dH \left({\bf u} \right)
&=& 
\frac{1}{2}\left( 
    \int_{[0,1]^2} \int_0^1 {\cal I}\left(u_1 \leq a^{\leftarrow}(u) \right)du dH(u_1,u_2) 
    +
    \int_{[0,1]^2} \int_0^1 {\cal I}\left(u_2 \leq a(u) \right)du dH(u_1,u2)
    \right)  \\
&& -     
\int_{[0,1]^2}  \int_0^1 {\cal I}\left(a^{\leftarrow}(u_1) \leq u, a(u_2) \leq u \right) du dH(u_1,u_2).
\end{eqnarray*}
The stated lemma can be deduced by applying Fubini's theorem on the three double integrals. 
\hfill $\square$


By Lemma \ref{lemma: H}, we obtain for $a_c(u) = F_2(c F^{\leftarrow}_1(u))$
$$
D_n(a_c)= (\phi(C_n))(a_c) 
\mbox{ and }
D(a_c)= (\phi(C))(a_c).
$$ 
this leads to 
$$
|| D_n(a_c)-D(a_c)||_{\infty} \leq 2 || C_n -C ||_{\infty}.  
$$ 
Classical results about empirical copulas gives uniform strong consistency, see Segers .... 
Similar arguments can be used for $\widehat{D}_n(\hat{a}_c)$. 
Now, we can consider the empirical process 
$$
\CB_n=\sqrt{n}(C_n-C),  \qquad \copula_n=\sqrt{n}(\widehat{C}_n-C).
$$
and we can write 

$$
\sqrt{n} \bigl( D_n(a_c)-D(a_c) \bigr) 
= (\phi(\CB_n))(a_c)
\mbox{ and } 
\sqrt{n} \bigl( \widehat{D}_n(\hat{a}_c)-D(\hat{a}_c) \bigr) = (\phi(\copula_n))(\hat{a}_c).
$$
We recall now that 
in the space $\ell^{\infty}([0,1]^d)$ equipped with the supremum norm, 
$\CB_n\conW\CB$, as $n\rightarrow\infty$, where $\CB$ is a C-Brownian bridge, and, as condition (S) holds, then $\copula_n\conW\copula$, as $n\rightarrow\infty$, 
where $\copula$ is the Gaussian process defined in \eqref{eq:cop_proc}, see \cite{seger12} for details.
In addition, $\hat{a}_c$ converges in probability to $a_c$.
The continuous mapping theorem then implies, as $n\rightarrow\infty$, 
$$
\sqrt{n} \bigl( D_n(a_c)-D(a_c) \bigr) =\phi(\CB_n)\conW\phi(\CB),
\quad
\sqrt{n} \bigl( \widehat{D}_n(\hat{a}_c)-D(\hat{a}_c) \bigr) = (\phi(\copula_n))
\conW\phi(\copula),
$$
in $\ell^{\infty}([0,1])$. 
From the continuity of its sample paths and by the form of the
covariance function \eqref{eq:covariance}, the Gaussian process $\copula$ satisfies
$$
\prob\{\forall\, u \in[0,1]: \copula(u,1)=\copula(1,u)=0\}=1.
$$
Ths provides all the elements to conclude the proposition. \hfill $\square$. 

\section{Expression of $D(c)$ in the bivariate GEV case}\label{sec: app D(c)}
 As $|a-b|= 2 \max(a,b)- a-b$, we have 
$$
2 D(c) = 2 \mathbb{E} \left[ \max \left( F_2\left(c Y_1\right), F_1 \left(  Y_2 /c\right) \right) \right]
-  \mathbb{E} \left[ F_2\left(c Y_1\right)\right] - \mathbb{E} \left[  F_1\left(  Y_2 /c\right) \right]
$$
To deal with each term, we recall that the quantile function of $ F(x; \xi,\sigma) =\exp\left[ -\left( \dfrac{x}{\sigma }\right)^{-1/\xi}\right]$
is $$ F^{-1}(u; \sigma,\xi) =  \sigma\left( -\log u \right)^{-\xi}  = \sigma z^{\xi} \mbox{, with } z = -1/\log(u), $$
This implies that 
$$
Y_i  \overset{\mathrm{d}}{=}  \sigma_i Z_i^{\xi_i},  
$$ 
where $Z_i$ follows an unit Fr\'echet.
If follows that, with 
$a_{12} =\left(\dfrac{c \sigma_1}{\sigma_2}\right)^{-1/\xi_2}$, 
$$
F_2\left(c Y_1\right) \overset{\mathrm{d}}{=}   \exp\left[ - \left(\dfrac{c Y_1}{ \sigma _2}\right)^{-1/\xi_2} \right] 
\overset{\mathrm{d}}{=} \exp \left( -a_{12} Z_1^{-\xi_1/\xi_2} \right) 
,
$$
then $$ F_2\left(c Y_1\right)\overset{\mathrm{d}}{=}  \exp \left( -a_{12} W_1 \right)  \mbox{ with } W_1= Z_1^{-\xi_1/\xi_2}.$$
In the same way, with $a_{21} = \left(\dfrac{ \sigma_2}{c\sigma_1}\right)^{-1/\xi_1}$,
$$
F_1\left(  Y_2 /c \right) \overset{\mathrm{d}}{=}  \exp\left[ -  \left(  \dfrac{Y_2 }{c\sigma_1}\right)^{-1/\xi_1} \right]  
\overset{\mathrm{d}}{=} \exp \left( -a_{21} Z_2^{-\xi_2/\xi_1} \right) $$

then $$
F_1\left(  Y_2 /c \right) \overset{\mathrm{d}}{=} \exp \left( -a_{21} W_2 \right)  \mbox{ with } W_2= Z_2^{-\xi_2/\xi_1}.
$$

By noticing that $W_i$ follows a Weibull distribution with $\PP(W_1 > w) =\exp(- w^{-\xi_2/\xi_1})$,  the  expectation  $\mathbb{E}[F_2\left(c Y_1\right)] $
can be linked as the Laplace transform of a Weibull r.v.
$$
\mathbb{E} \left[F_2\left(c Y_1\right)\right] = \mathbb{E} \left[ \exp \left( -a_{12} W_1 \right) \right]
\mbox{ and } 
 \mathbb{E}  \left[ F_1\left( Y_2 /c\right)\right] = \mathbb{E}\left[ \exp \left( -a_{21} W_2 \right) \right].
$$
For the bivariate structure, we can write that, for any $u\in (0,1)$, 
\begin{eqnarray*}
\mathbb{P} \left[ \max \left( F_2\left(c Y_1\right), F_1\left(  Y_2 /c\right) \right) \leq u \right] 
&=&  \mathbb{P} \left[ \max \left( \exp \left( -a_{12} Z_1^{-\xi_1/\xi_2} \right),  \exp \left( -a_{21} Z_2^{-\xi_2/\xi_1} \right) \right) \leq u \right],\\
&=&  \mathbb{P} \left[ Z_1  \leq \left( \frac{-a_{12}}{\log u} \right)^{\xi_2/\xi_1},\,Z_2  \leq \left( \frac{-a_{21}}{\log u} \right)^{\xi_1/\xi_2}\right],\\
&=&  \exp \left\{ -V\left[ \left(\frac{-a_{12}}{\log u} \right)^{\xi_2/\xi_1} ,\left( \frac{-a_{21}}{\log u} \right)^{\xi_1/\xi_2} \right]\right\}.\\
\end{eqnarray*}

Since the r.v. $\max\left(F_2\left(c Y_1\right), F_1\left(  Y_2 /c\right)\right) \leq u$ is positive, in the general setup, we have 
\begin{eqnarray}\label{eq: D bi-GEV}
D &=& \int_0^1  \left( 1- \exp \left\{  -V\left[ \left( \frac{a_{12}}{-\log u} \right)^{\xi_2/\xi_1} ,\left( \frac{a_{21}}{-\log u} \right)^{\xi_1/\xi_2} \right]\right\}  \right) du \nonumber \\
&&- \frac{1}{2}  \mathbb{E} \left[ \exp \left( -a_{12} W_1 \right) \right] 
- \frac{1}{2}  \mathbb{E} \left[\exp \left( -a_{21} W_2 \right) \right],
\end{eqnarray}
where $W_i$ follows a Weibull distribution with $\PP(W_1 > w) =\exp(- w^{\xi_1/\xi_2})$. 
Note that 
\begin{equation*}
    \left(a_{12}\right)^{\frac{\xi_2}{\xi_1}}= \dfrac{1}{a_{21}}  
\end{equation*}

Conversely, $\left(a_{21}\right)^{\frac{\xi_1}{\xi_2}}=\dfrac{1}{a_{12}}$\\

\section{Homogeneous case}\label{sec: append homog} 

In the special case where 
\textbf{$\xi_1=\xi_2=\xi$}, we denote 
$\theta_c = V\left(a_{12},a_{21}\right)$, where 
$a_{12}=\left(\dfrac{c\sigma_1}{\sigma_2}\right)^{-1/\xi} = 1/a_{21}$. 
Then, we have
\begin{eqnarray*}
\mathbb{P} \left[ \max \left( F_2\left(c Y_1\right), F_1\left(  Y_2 /c\right) \right) \leq u \right] 
&=&  \exp \left\{ V\left[ \left(\dfrac{\sigma_2}{c\sigma_1}\right)^{-1/\xi} , \left(\dfrac{c\sigma_1}{\sigma_2}\right)^{-1/\xi}\right] \log u \right\}, \\
&=& u^{V\left( a_{12} ,a_{21} \right)}.\\
\end{eqnarray*}
We can write
\begin{equation}
    D = \int_0^1  1- u^{\theta_c} du
- \frac{1}{2}  \mathbb{E}\left[ \exp \left( -a_{12} W_1 \right) \right]  
- \frac{1}{2}  \mathbb{E} \left[\exp \left( -a_{21} W_2 \right) \right]  
\end{equation}
where $W_i, i=1,2$ has cdf equal to $\exp(-x).$\\
Hence, 
\begin{equation*}
    D = \dfrac{\theta_c}{\theta_c +1} - \dfrac{1}{2(1+a_{12})}-\dfrac{1}{2(1+a_{21})}.
\end{equation*}
To minimise $D$ as a function of $c$, we study the variations of $r: x \longmapsto \dfrac{V\left(x, \frac{1}{x}\right)}{1 + V\left(x, \frac{1}{x}\right)}
- \dfrac{1}{2(1+x)}-\dfrac{x}{2(1+x)}.$ We suppose that \textbf{$V$ is differentiable.} If the previous function $r$ admits a minimum, its derivative cancels in some $c_0$. The $r'$ cancels if and only if the derivative of $x \longmapsto \dfrac{V\left(x, \frac{1}{x}\right)}{1+V\left(x, \frac{1}{x}\right)}$ cancels, if and only if there exists $x$ s.t. $\dfrac{\partial V}{\partial x}\left(x, \frac{1}{x}\right) = \dfrac{1}{x^2}\dfrac{\partial V}{\partial y}\left(x, \frac{1}{x}\right).$
In the special case where the \textbf{dependence is logistic} {\it i.e.} $$V(x,y) = \left(\dfrac{1}{x^{1/\alpha}}+ \dfrac{1}{y^{1/\alpha}}\right)^{\alpha},$$
we have $\dfrac{\partial V}{\partial x}\left(x, \frac{1}{x}\right) =\dfrac{\partial V}{\partial y}\left(x, \frac{1}{x}\right), \quad\text{ for all positive } x$. Therefore, if $r$ admits a minimum, it is for $x=\pm 1.$ Eventually, for logistic dependence, $D$ is minimal for $$c=\dfrac{\sigma_2}{\sigma_1}.$$

\end{appendix}

\bibliographystyle{rss}
\bibliography{biblio}
\end{document}